\documentclass[%
 reprint,
 amsmath,amssymb,
 aps,
]{revtex4-1}

\usepackage{graphicx}
\usepackage{dcolumn}
\usepackage{bm}
\usepackage[version=4]{mhchem}
\usepackage{braket}
\usepackage{amsmath}
\usepackage{bm}

\newcommand{\ketbra}[2]{\ket{#1}\!\bra{#2}}

\usepackage{xcolor}

\begin{document}


\title{Observability Architecture for Quantum-Centric Supercomputing Workflows}

\author{Naoki Kanazawa}
\thanks{Equal contributions}
\thanks{Now at another institution.}
\affiliation{%
 IBM Quantum, IBM Research – Tokyo, Tokyo 103-8510, Japan
}%

\author{Yuto Morohoshi}
\email{u542400c@ecs.osaka-u.ac.jp}
\thanks{Equal contributions}
\affiliation{%
IBM Quantum, IBM Research – Tokyo, Tokyo 103-8510, Japan
}%
\affiliation{%
Graduate School of Engineering Science, The University of Osaka, 1-3 Machikaneyama, Toyonaka, Osaka 560-8531, Japan
}%

\author{Hitomi Takahashi}
\affiliation{%
 IBM Quantum, IBM Research – Tokyo, Tokyo 103-8510, Japan
}%

\author{Yukio Kawashima}
\affiliation{%
 IBM Quantum, IBM Research – Tokyo, Tokyo 103-8510, Japan
}%

\author{Hiroshi Horii}
\email{horii@jp.ibm.com}
\affiliation{%
 IBM Quantum, IBM Research – Tokyo, Tokyo 103-8510, Japan
}%

\author{Kengo Nakajima}
\email{nakajima@cc.u-tokyo.ac.jp}
\affiliation{%
 Information Technology Center, University of Tokyo, Kashiwa, Chiba 277-8589, Japan 
}%

\date{\today}

\begin{abstract}
Quantum-centric supercomputing (QCSC) workflows often involve hybrid classical–quantum algorithms that are inherently probabilistic and executed on remote quantum hardware, making them difficult to interpret and limiting the ability to monitor runtime performance and behavior. The high cost of quantum circuit execution and large-scale high-performance computing (HPC) infrastructure further restricts the number of feasible trials, making comprehensive evaluation of execution results essential for iterative development. We propose an observability architecture tailored for QCSC workflows that decouples telemetry collection from workload execution, enabling persistent monitoring across system and algorithmic layers and retaining detailed execution data for reproducible and retrospective analysis, eliminating redundant runs. Applied to a representative workflow involving sample-based quantum diagonalization, our system reveals solver behavior across multiple iterations. This approach enhances transparency and reproducibility in QCSC environments, supporting infrastructure-aware algorithm design and systematic experimentation.
\end{abstract}

\maketitle


\section{\label{sec:intro}Introduction}

Quantum-centric supercomputing (QCSC) is emerging as a promising paradigm for addressing computational problems that remain intractable on classical systems alone. Recent studies have examined the architectural and algorithmic foundations of QCSC, highlighting its potential to accelerate scientific discovery through hybrid quantum–classical workflows~\cite{lanes2025frameworkquantumadvantage, doi:10.1021/acs.jctc.5c00075, 75pv-hbrx, kanno2023quantum, 10.3389/fcomp.2025.1528985}. These workflows integrate quantum kernels within classical orchestration and are typically executed on high-performance computing (HPC) infrastructure.

However, integrating quantum and classical components introduces new challenges in scalability, interpretability, and resource management~\cite{Beck_2024, bacher2025quantumresourcesresourcemanagement, mantha2025pilotquantumquantumhpcmiddlewareresource, wennersteen2025towards, mansfield2025practicalexperiencesintegratingquantum}. A defining characteristic of QCSC workflows is their reliance on algorithms that combine probabilistic quantum subroutines with classical control logic, particularly at HPC scale~\cite{ANTUNES2024100655}. These algorithms are often sensitive to hyperparameter configurations, hardware noise, and execution timing, while limited visibility into remote execution further complicates understanding their behavior, making algorithmic dynamics hard to predict and reproduce ~\cite{Tilly_2022}.

The complexity is further compounded by the high cost of execution, which includes not only quantum circuit sampling but also the overhead of HPC job scheduling and resource reservation. These constraints limit the number of feasible trials and underscore the need for detailed monitoring and post-hoc or even in-process data analysis.

While observability has been extensively studied in HPC environments, existing approaches predominantly focus on system-centric metrics—for example, CPU utilization, memory bandwidth, and thermal states~\cite{8049016, 9229636, balis2024observabilityscientificapplications, 10.1145/3673038.3673100}. These metrics are primarily consumed by system administrators for performance tuning and anomaly detection, offering limited insight into application-level behavior. In contrast, QCSC workflows demand application-centric observability—encompassing the ability to monitor solver behavior, track intermediate artifacts, and analyze scientific outcomes.

Prior efforts in quantum computing have introduced hardware-level observability using sensor networks to infer error characteristics~\cite{11018264}; however, these approaches are limited to low-level signals and fail to address the complexity of hybrid algorithm execution across HPC and quantum resources.

In this study, we propose an observability architecture designed to integrate seamlessly with workflow management systems and QCSC execution environments, enabling persistent monitoring, structured storage, and retrospective analysis of QCSC workloads. This integration ensures that observability components operate in concert with workflow orchestration tools—such as Prefect—while interfacing with HPC and quantum platforms, thereby providing application-level insights without disrupting primary workload execution.

The key contributions of this work are as follows:
\begin{itemize}
    \item \textbf{Conceptual Framework:} We formalize application-level observability for QCSC workflows through a workflow metrics pyramid, distinguishing system-centric and domain-centric telemetry layers.
    \item \textbf{Architecture Design:} We propose a decoupled observability architecture that integrates with workflow orchestration systems to enable persistent monitoring and post-hoc analysis without interfering with primary execution.
    \item \textbf{Reference Implementation:} We implement the proposed architecture on the Miyabi supercomputer and IBM Quantum systems, leveraging Prefect for orchestration and Apache Superset for visualization.
    \item \textbf{Practical Demonstration:} We validate the framework through a closed-loop SQD workflow using differential evolution, illustrating how telemetry informs algorithmic behavior and resource efficiency.
\end{itemize}

The remainder of the paper is organized as follows. Section~\ref{sec:arch} presents a generic system architecture for QCSC workload execution. Section~\ref{sec:impl} describes a reference implementation on the Miyabi supercomputer and IBM Quantum. Section~\ref{sec:usecase} demonstrates the utility of the observability framework using a practical example involving the differential evolution algorithm applied to sample-based quantum diagonalization (SQD). We conclude with a discussion of future directions in Section~\ref{sec:conclusion}.

\begin{figure}[tb]
\centering
\includegraphics[width=0.9\columnwidth]{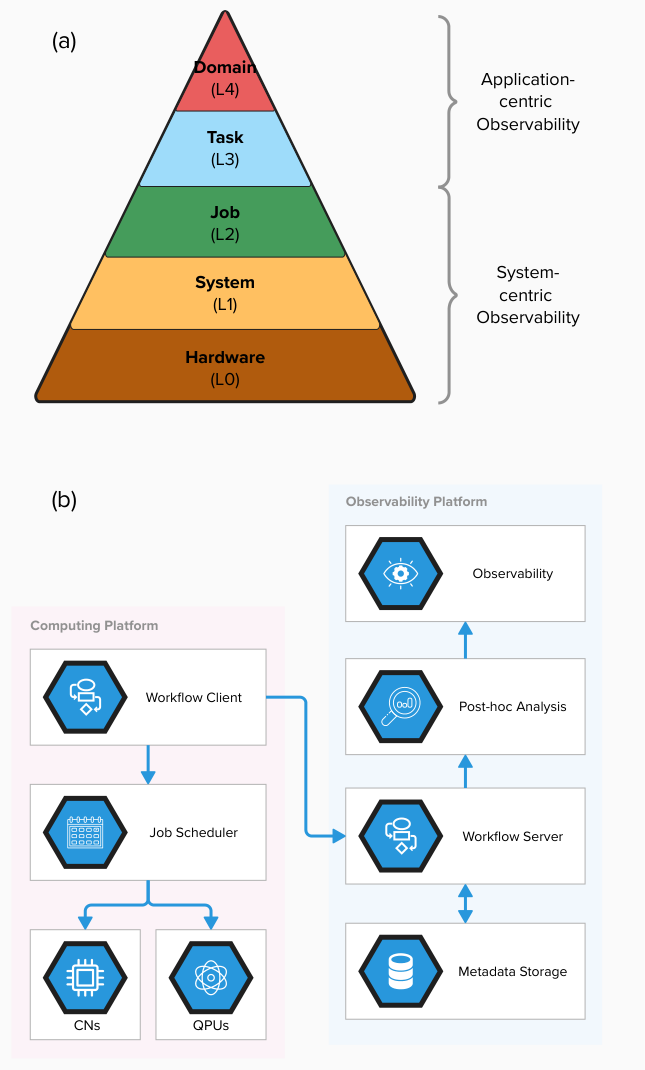}
\caption{
(a) Workflow metrics pyramid showing categories of telemetry data and (b) schematic diagram of the QCSC observability architecture.
}
\label{fig:overview}
\end{figure}

\section{\label{sec:arch}Architecture}

To structure our discussion, we introduce a workflow metrics pyramid that categorizes telemetry data into five levels, as shown in Fig.~\ref{fig:overview}(a):

\begin{enumerate}
    \item[L0.] Hardware-level: power, thermal, ...
    \item[L1.] System-level: CPU, memory, I/O, network ...
    \item[L2.] Job-level: accounting, node usage, ...
    \item[L3.] Task-level: artifacts, wall-clock times, ...
    \item[L4.] Domain-level: solver convergence, fidelity, ...
\end{enumerate}

While L0–L2 represent system-centric observability, L3–L4 are essential for understanding algorithmic behavior and guiding experimental design. These higher-level metrics are not accessible through job scheduler APIs alone and require tight integration with workflow orchestration tools.

Figure~\ref{fig:overview}(b) presents a schematic overview of the proposed architecture, illustrating the separation between the computing platform and the observability platform. This separation is a deliberate design choice, motivated by both performance considerations and architectural clarity.

In QCSC workflows, merely aggregating all available data is insufficient to understand solver behavior. Instead, we isolate the observability platform to enable asynchronous processing of raw workflow data. This allows the extraction of high-level, interpretable metrics without interfering with the execution environment.

The observability platform operates as a persistent service, connected to the computing platform via a network. It receives metadata from the workflow client, stores it in a structured database, and supports retrospective analysis of solver behavior. This separation is particularly important given the computational cost of analysis and the need to preserve the reliability of the primary workload.

The computing platform follows a conventional HPC architecture, where a workflow client orchestrates tasks submitted to a job scheduler and dispatched to compute resources, including classical compute nodes (CNs) and quantum processing units (QPUs).

\section{\label{sec:impl}Reference Implementation}

\begin{figure*}[tb]
\centering
\includegraphics[width=0.8\textwidth]{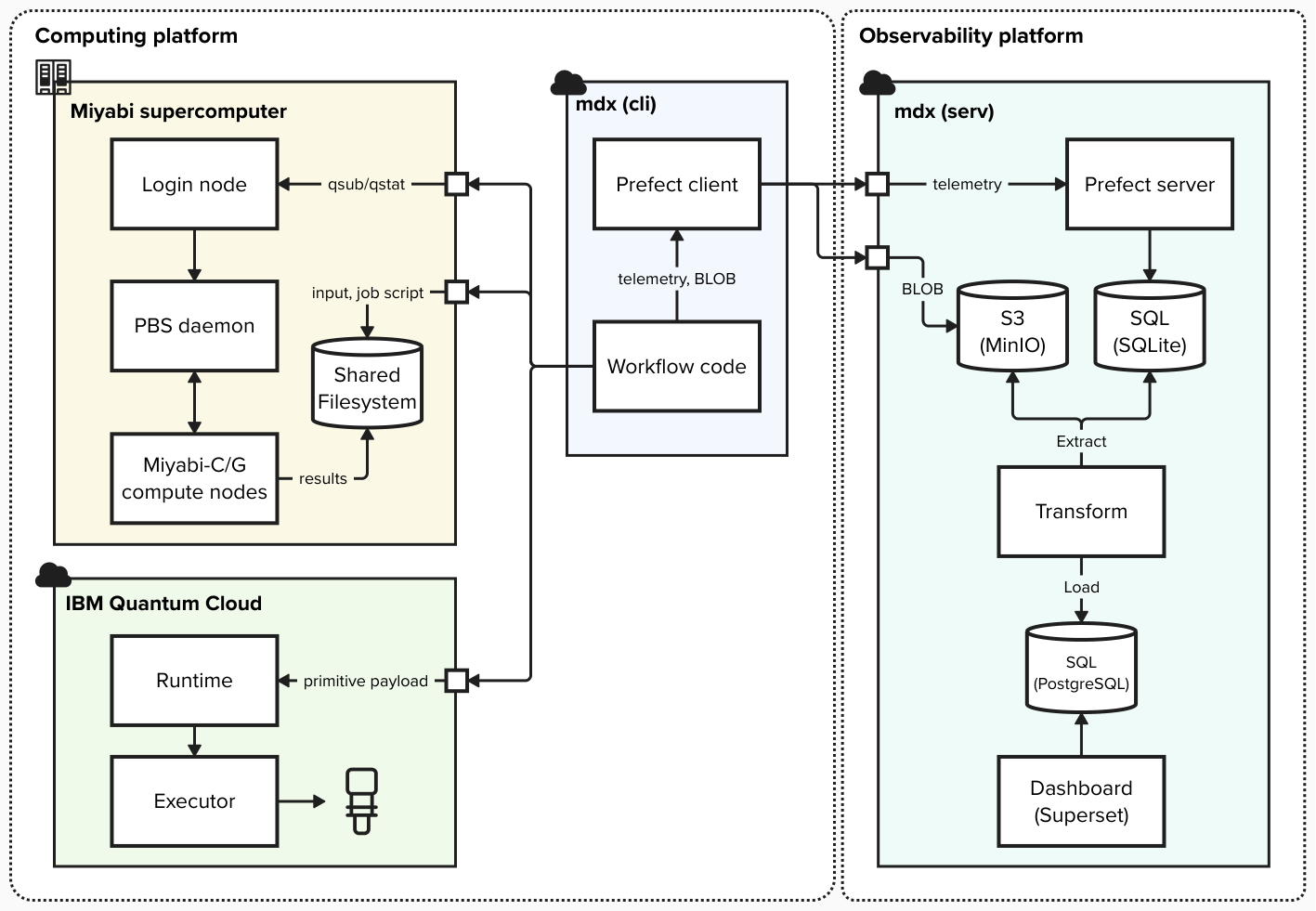}
\caption{
Component diagram of QCSC observability architecture implemented on the Miyabi supercomputer and mdx platform.
}
\label{fig:component_miyabi_mdx}
\end{figure*}

We implemented our QCSC observability architecture on the Miyabi supercomputer, jointly operated by the University of Tokyo and University of Tsukuba. The system uses the Altair PBS Professional job scheduler for resource management. Additionally, we leverage the mdx platform~\cite{9927975} to host two virtual machines: one for the computing platform and another for the observability platform.

Figure~\ref{fig:component_miyabi_mdx} illustrates the component architecture. The \texttt{cli} machine serves as the workflow client, hosting the application code and interfacing with both the Qiskit Runtime REST API and the Miyabi login node for quantum and HPC workload execution, respectively. Secure SSH is used to issue PBS commands (\texttt{qsub}, \texttt{qstat}), and the Miyabi remote file system is mounted on the \texttt{cli} machine via SFTP. This setup allows the workflow to generate PBS job scripts and input data directly on the Miyabi file system.

This loosely coupled execution model enables elastic job scheduling, albeit at the cost of frequent queuing. However, it avoids algorithmic performance bottlenecks described by Amdahl’s law, as quantum primitive executions typically do not benefit from parallel classical computation.

The \texttt{serv} machine hosts the workflow server, which collects telemetry from the \texttt{cli} machine and provides an observability dashboard. The two machines communicate over the public internet with access control.

We use Prefect as the workflow orchestrator and Apache Superset to build the observability dashboard. A Prefect client on the \texttt{cli} machine manages workflow execution and collects telemetry during task execution. For instance:
\begin{itemize}
    \item L2 telemetry (e.g., Qiskit Runtime and PBS job metrics) is automatically extracted by a custom job executor wrapper and formatted into Prefect table artifacts for resource allocation analysis.
    \item L3 telemetry (e.g., wall-clock times) is recorded as execution metadata by the Prefect client, aiding in performance bottleneck identification.
    \item L4 telemetry (e.g., workflow-specific intermediate data) is captured directly by the user-defined workflow code.
\end{itemize}
The Prefect server on the \texttt{serv} machine stores telemetry in its built-in SQL database and exposes a REST API for accessing artifacts and metadata. Additionally, the \texttt{serv} machine hosts an S3-compatible object store for binary large objects (e.g., raw bitstrings from quantum samplers) that are unsuitable for relational databases.

Post-hoc analysis is performed via an ETL (Extract, Transform, Load) pipeline implemented as a server-side Prefect workflow. 
Importantly, the telemetry data collected from the workflow executions are persistently stored in both database. This design enables the incorporation of new metrics without rerunning experiments, requiring only modifications to the subsequent ETL steps. The normalized data are then stored in a PostgreSQL backend and visualized using Apache Superset.
This pipeline enables in-situ introspection of the behavior of the QCSC algorithms.

\section{\label{sec:usecase}Use Case}
In~\cite{Robledo_Moreno_2025}, the authors introduced the SQD technique to simulate the ground state energy of a chemistry Hamiltonian, which is inspired by the quantum-selected configuration interaction technique \cite{kanno2023quantum}. Electronic configurations are sampled as bitstrings $\tilde{\chi}$ using a noisy quantum primitive. These sampled configurations $\tilde{\chi}$ span a subspace, from which an approximate ground state energy is computed using the Davidson diagonalization method on multiple classical compute nodes. The target Hamiltonian is mapped to a quantum circuit using the local unitary cluster Jastrow (LUCJ) ansatz~\cite{D3SC02516K}. Circuits are initialized with classically computed coupled cluster singles and doubles (CCSD) amplitudes.

The authors in~\cite{shirakawa2025closedloopcalculationselectronicstructure} further optimized the LUCJ circuit parameters using the differential evolution (DE) algorithm~\cite{10.1023/A:1008202821328}. While this experiment demonstrated improvements in solution quality, the impact of stochastic quantum kernels in such heuristic algorithms remains an open question. We find this to be a compelling problem setting to demonstrate the capabilities of our QCSC observability architecture.

\subsection{Setup}

\begin{figure}[t!]
\centering
\includegraphics[width=1.0\columnwidth]{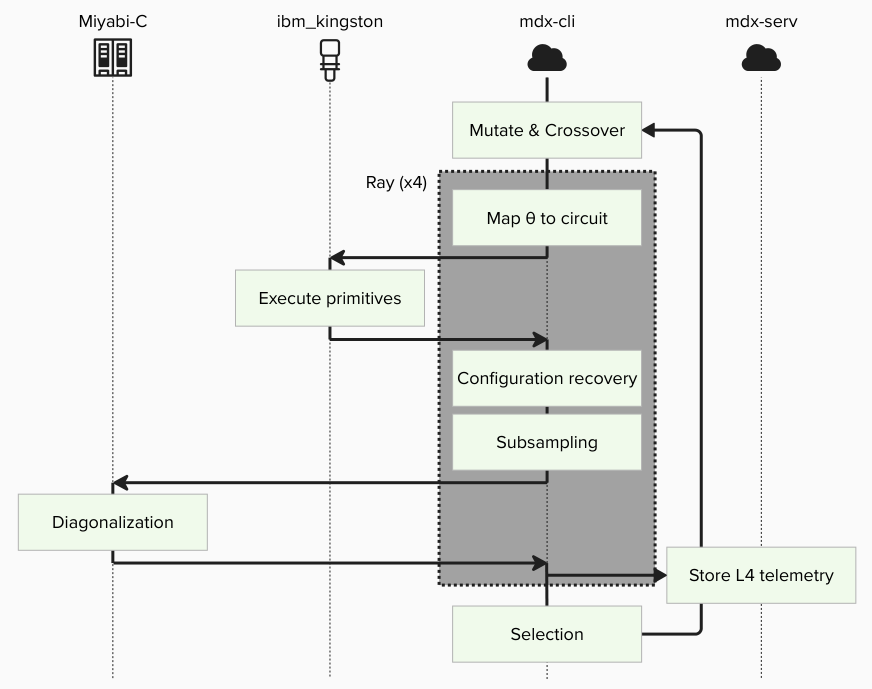}
\caption{
Timing diagram of the differential evolution algorithm workflow.
The shaded box on the \texttt{cli} machine is implemented as a single task and executed in parallel for all populations using the Ray executor integration \texttt{prefect-ray}.
Important SQD variables in Table~\ref{tab:sqd-telemetry} are stored on the \texttt{serv} machine for post-hoc ETL pipeline execution.
}
\label{fig:timing-diagram}
\end{figure}

In our experiment, we consider the ground state energy of the synthetic \ce{[Fe4S4(SHC3)4]^{2-}} cluster, abbreviated as [4Fe-4S], as reported in~\cite{Robledo_Moreno_2025}. We use 64 CPU nodes from Miyabi-C for subspace diagonalization and the IBM Heron R2 processor \texttt{ibm\_kingston} to sample bitstrings with LUCJ circuits. The closed-loop SQD workflow is illustrated in Fig.~\ref{fig:timing-diagram}.

The chemistry Hamiltonian of [4Fe-4S] is mapped to quantum circuits as
\begin{equation}
\ket{\Psi}_i = \mathcal{U}_\mathrm{LUCJ}(\boldsymbol{\theta}_{i,g}) \ket{\bm{\mathrm{x}}_\text{RHF}},
\end{equation}
where $\mathcal{U}_\mathrm{LUCJ}$ is the unitary evolution under a truncated LUCJ circuit in $L$-product form, $\boldsymbol{\theta}_{i,g}$ are the circuit parameters for the $i$-th population at iteration (generation) $g$ in the differential evolution loop, and $\bm{\mathrm{x}}_\text{RHF}$ is the bitstring representing the restricted Hartree–Fock (RHF) state.
We use a truncated LUCJ circuit with $L=2$, resulting in approximately 6.3k two-qubit gates after SABRE layout optimization~\cite{zou2024lightsabrelightweightenhancedsabre} with layout trials $10^5$. 

In each optimization iteration, trial populations $\boldsymbol{\tilde{\theta}}_{i,g}$ are generated via mutation and crossover. Mutant vectors $\boldsymbol{v}_{i,g}$ are formed by using two vector differences:
\begin{equation}
\boldsymbol{v}_{i,g} = \boldsymbol{\theta}_{\text{best},g} + F\left(\boldsymbol{\theta}_{i0,g} - \boldsymbol{\theta}_{i1,g} + \boldsymbol{\theta}_{i2,g} - \boldsymbol{\theta}_{i3,g}\right),
\label{eq:mutant}
\end{equation}
where $i0, i1, i2, i3$ are distinct random indices and $F$ is a scalar DE parameter. The trial population $\boldsymbol{\tilde{\theta}}_{i,g}$ is then formed as
\begin{equation}
\tilde{\theta}_{i,g,j} =
\begin{cases}
v_{i,g,j}, & \text{if } r \leq C_r \text{ or } j = j_\text{rand}, \\
\theta_{i,g,j}, & \text{otherwise},
\end{cases}
\end{equation}
where $j$ indexes the parameter vector, $j_\text{rand}$ ensures that at least one element is taken from the mutant vector, and $C_r \in (0, 1]$ is another DE parameter. We use $F = 0.6$ and $C_r = 0.9$ in this experiment.

Quantum circuits for all populations are executed with trial populations $\boldsymbol{\tilde{\theta}}_{i,g}$ and measured in the computational basis. Repeating this yields measurement outcomes $\tilde{\chi}_i$ as bitstrings $\bm{\mathrm{x}} \in \{0,1\}^M$, representing electronic configurations. 
The TZP-DKH basis set is used to describe the [4Fe–4S] system, and 3d orbitals of Fe atoms, 3p orbitals of S atoms, and Fe-SCH3 ligand orbitals are used to construct the active space. This results to producing bitstrings of length $M=72$, corresponding to the number of spin orbitals using Jordan–Wigner transformation.
To mitigate circuit noise, we apply dynamical decoupling using the XY-4 sequence~\cite{Ezzell_2023} and a reset-mitigation scheme that post-selects outcomes based on measurements of prepared initial states. Each execution produces $500 \cdot 10^3$ samples, with an average shot retention rate of approximately 0.469 after post-selection.

From the noisy configurations $\tilde{\chi}_i$, we compute the ground state energy of [4Fe-4S] using the conventional SQD procedure. Unlike approaches that refine SQD iteratively on classical hardware, our method performs the entire SQD step once per iteration, as part of the outer differential evolution loop.
First, we generate a partially recovered noiseless configuration $\chi_{R,i}$ using a self-consistent recovery scheme. 
To facilitate conservation of total spin, we create unique configurations $\chi_{R,i}^\text{u}$ of length $M/2=36$ for up- and down-spin bitstrings as presented in the original work~\cite{shirakawa2025closedloopcalculationselectronicstructure}.
We consider the Hamiltonian projected onto the $d = 10^8$-dimensional space defined by the subspace $\mathcal{S}$:
\begin{equation}
\hat{H}_{\mathcal{S},i} = \sum_{\bm{\mathrm{x}}_i \in \mathcal{S}_i + \chi_\text{co}} \ketbra{\bm{\mathrm{x}}_i}{\bm{\mathrm{x}}_i},
\end{equation}
where $\mathcal{S}_i$ is constructed from bitstrings subsampled from $\chi_{R,i}^\text{u}$, and $\chi_\text{co}$ represents carryover bitstrings from the previous DE iteration.
The ground state energy $E_i$ of $\hat{H}_{\mathcal{S},i}$ is computed using the Davidson method on Miyabi-C. Carryover bitstrings $\chi_{\text{co},i}$ are selected based on the computed ground state wavefunction.

Finally, the next-iteration population $\boldsymbol{\theta}_{i,g+1}$ is selected based on the trial energy:
\begin{equation}
\boldsymbol{\theta}_{i,g+1} =
\begin{cases}
\boldsymbol{\tilde{\theta}}_{i,g}, & \text{if } E_{i,g} \leq E_{i,g-1}, \\
\boldsymbol{\theta}_{i,g}, & \text{otherwise}.
\end{cases}
\end{equation}
The best carryover $\chi_{\text{co, best}}$ is selected from the population with the lowest energy and used in the next SQD iteration for all populations.

We perform 20 iterations with 4 populations, resulting in 80 quantum primitive executions and subspace diagonalizations on Miyabi-C. During execution, intermediate data listed in Table~\ref{tab:sqd-telemetry} is stored in object storage in compressed NumPy \texttt{.npz} format.

\begin{table*}[t]
\centering
\begin{tabular}{p{0.2\textwidth}|p{0.6\textwidth}}
\hline
\texttt{ucj\_parameter} & Trial LUCJ circuit parameters $\boldsymbol{\tilde{\theta}}$ \\
\texttt{raw\_bitstrings} & Sampled bitstrings before configuration recovery $\tilde{\chi}$ \\
\texttt{recovered\_bitstrings} & Sampled bitstrings after configuration recovery $\chi_R$ \\
\texttt{alphadets} & Up-spin bitstrings span a subspace Hamiltonian $\hat{H}_\mathcal{S}$ \\
\texttt{avg\_occupancy} & Average occupancy of spin orbitals at computed ground state. \\
\texttt{carryover} & Carryover bitstrings $\chi_{\text{co}}$ \\
\hline
\end{tabular}
\caption{L4 telemetry in the SQD trial}
\label{tab:sqd-telemetry}
\end{table*}

\begin{figure*}[p!]
\centering
\includegraphics[width=1.0\textwidth, height=0.8\textheight, keepaspectratio]{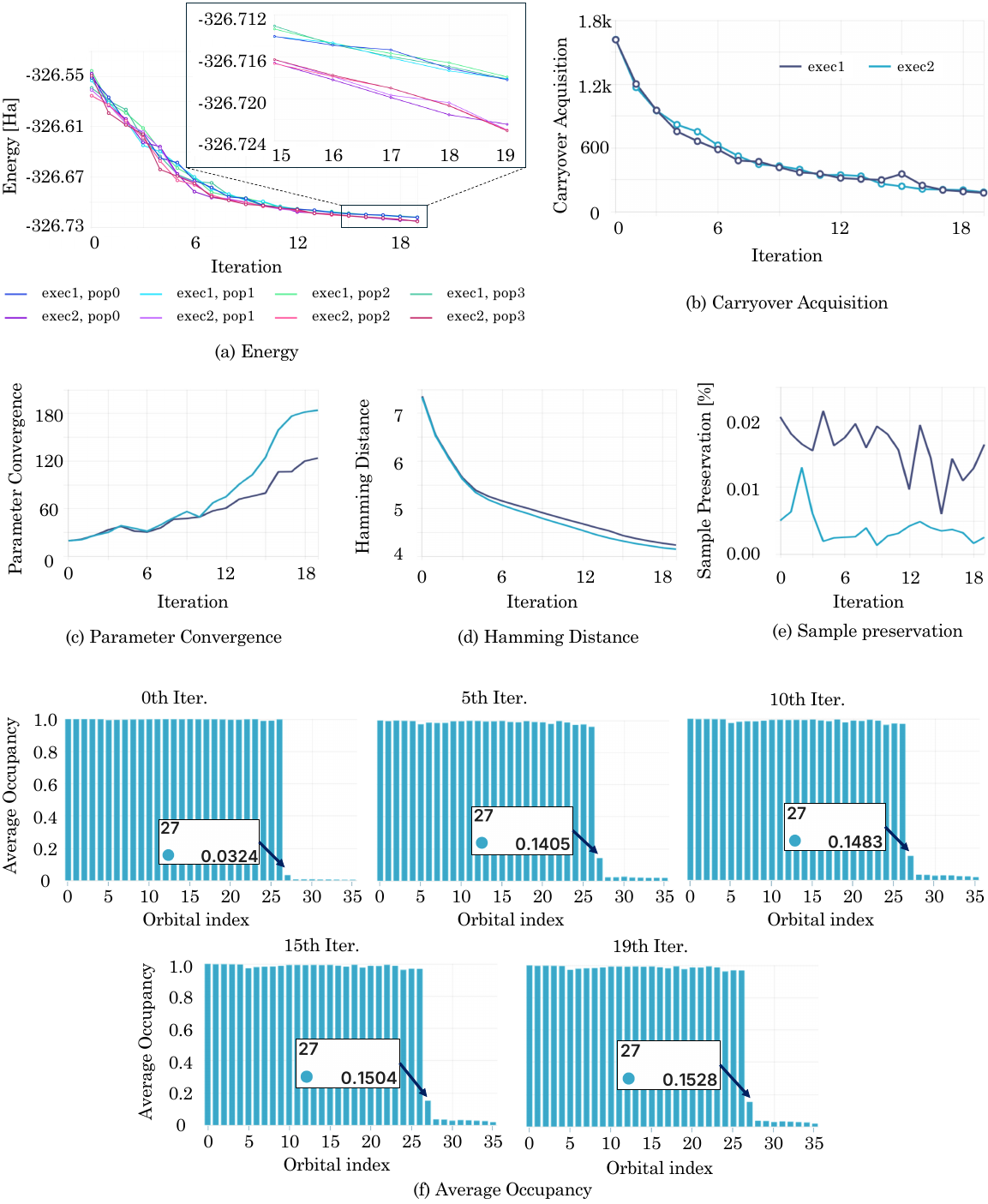}
\caption{
Observability dashboard of domain-level (L4) metrics in the closed-loop SQD workflow. All panels except (f) show trends of various metrics toward an increase in DE iterations from two independent executions under the identical setup.
The panels (f) show the average occupancy of each spatial orbital at different iterations of the first execution. The legend indicates the average occupancy of the 27th orbital, corresponding to the first unoccupied orbital in the RHF reference.
See the main text for the details of each metric.
}
\label{fig:dashboard-sqd}
\end{figure*}

\begin{figure*}[p!]
\centering
\includegraphics[width=1.0\textwidth, height=0.8\textheight, keepaspectratio]{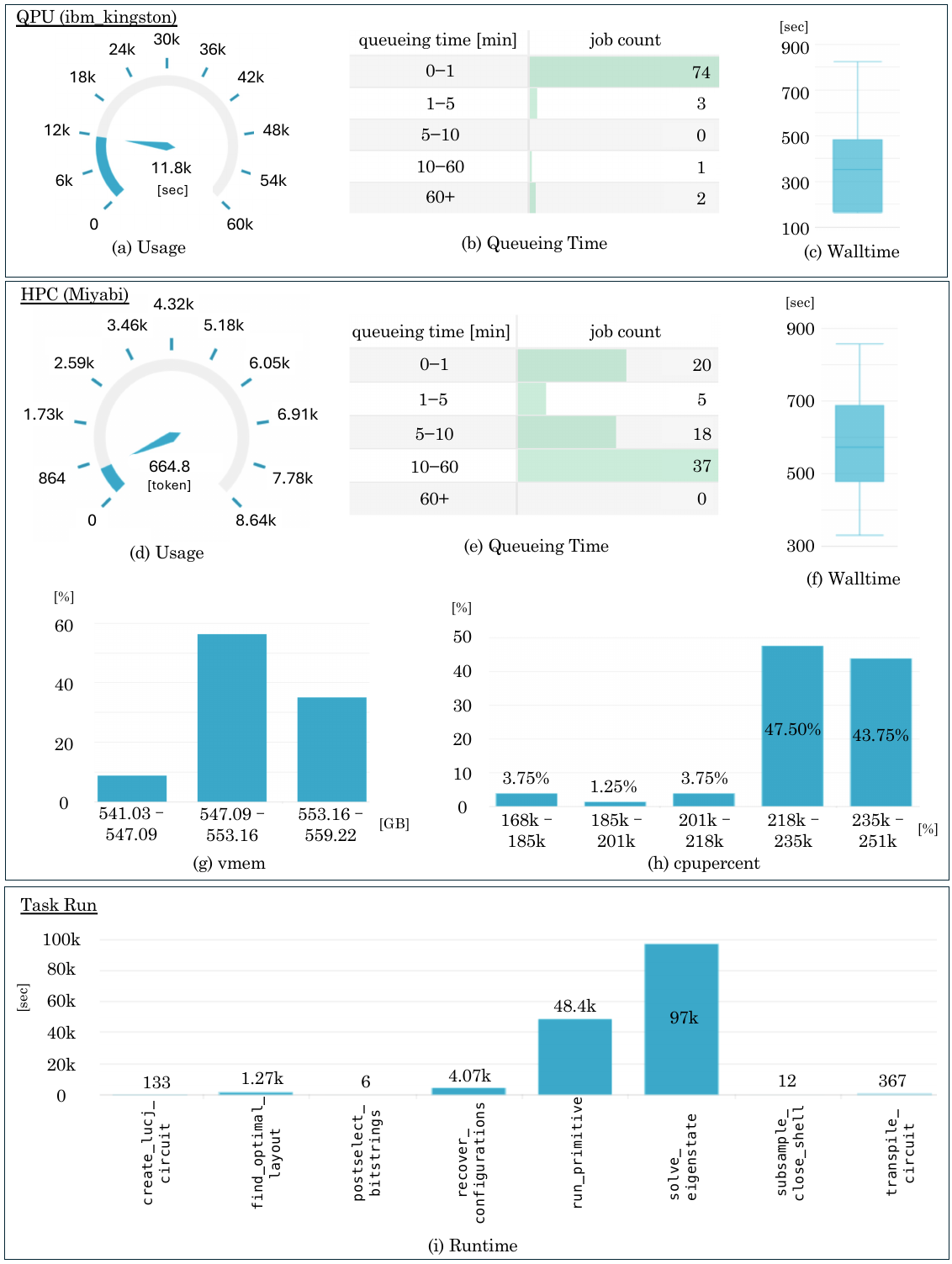}
\caption{
Observability dashboard of performance metrics (L3 and L2) from the first execution of the closed-loop SQD workflow.
The top and middle box show statistical data of the L2 telemetries for 80 jobs in the QPU and HPC, respectively.
The histograms represent performance distributions for these jobs.
HPC job metrics are collected by the \texttt{qstat -f} command.
Refer to the PBS reference guide \cite{altair_pbs_reference_2021} for the definitions.
(a) QPU usage relative to the allocated time limit of 60,000 seconds.
(b) Queueing time for QPU jobs.
(c) QPU wall-clock time, including primitive payload compilation, execution, and post-processing.
(d) HPC usage relative to the allocated token limit of 8,640 tokens.
(e) Queueing time for HPC jobs, computed by \texttt{stime} - \texttt{etime}.
(f) HPC wall-clock time, indicated by \texttt{walltime}.
(g) Total virtual memory allocated across all concurrent processes within a job, indicated by \texttt{resources\_used.vmem}.
(h) Maximum CPU utilization rate, indicated by \texttt{resources\_used.cpupercent}.
(i) Execution time for dominant Prefect tasks.
}
\label{fig:dashboard-performance}
\end{figure*}

\subsection{Metrics}

We define several domain-level metrics derived from the telemetry listed in Table~\ref{tab:sqd-telemetry}. These metrics are designed to capture algorithmic behavior and provide interpretable signals for workflow analysis:

\begin{description}
\item[Carryover acquisition] 
This metric measures the number of new carryover bitstrings introduced at iteration $g$. Let $\chi_{\text{co}, g}$ denote the set of carryover bitstrings at iteration $g$. The acquisition is given by:
\begin{equation}
|\chi_{\text{co}, g} \setminus \chi_{\text{co}, g-1}| \quad \text{for } g > 0,
\end{equation}
where $\setminus$ denotes set difference and $|\cdot|$ represents cardinality. In our workflow, a single set of carryover bitstrings with the lowest energy estimate among populations is selected per iteration.

\item[Parameter convergence] 
This metric quantifies the similarity among trial populations in the DE algorithm. It is computed as the mean pairwise Euclidean distance:
\begin{equation}
\frac{1}{C}\sum_{i<j} d(\tilde{\boldsymbol{\theta}}_{i,g}, \tilde{\boldsymbol{\theta}}_{j,g}),
\end{equation}
where $d(\cdot,\cdot)$ denotes Euclidean distance and $C = \binom{4}{2} = 6$ for four populations in this study. A decreasing trend indicates convergence toward similar parameter values.

\item[Hamming distance] 
This metric computes the average Hamming distance between each carryover bitstring and the RHF reference state. For a closed-shell system with $N_e$ electrons and $M/2$ spin orbitals, the reference represents the restricted Hartree–Fock configuration: the lowest $N_e$ spin orbitals are occupied, and the remaining $(M/2 - N_e)$ orbitals are unoccupied. In binary form, this appears as a string with $N_e$ ones and $(M/2 - N_e)$ zeros, such as \texttt{000...0111...1} in our implementation. 
This metric provides insight into how far the optimized configuration deviates from the RHF baseline, which reflects the degree of electron correlation captured by the workflow.

\item[Sample preservation]
This metric measures the fraction of bitstrings retained after configuration recovery:
\begin{equation}
\frac{|\tilde{\chi} \cap \chi_R|}{|\chi_R|},
\end{equation}
where $\tilde{\chi}$ is the raw sampled set and $\chi_R$ is the recovered set. A low ratio indicates strong influence of noise and recovery procedures on the final configuration distribution.
\end{description}

\subsection{Workflow behavior analysis}

We executed the closed-loop SQD workflow twice under identical conditions to evaluate reproducibility and algorithmic dynamics. Figure~\ref{fig:dashboard-sqd}(a) shows that the minimum ground-state energy in the first execution was $-326.71819$ Ha (population 0), while the second execution achieved $-326.72307$ Ha (population 3). Despite fixing all accessible random seeds, this discrepancy is most likely due to the inherent nondeterminism of quantum measurements.

Figure~\ref{fig:dashboard-sqd}(b) illustrates the trend in carryover acquisition, which decreases as optimization progresses. This behavior suggests that determinants contributing significantly to energy reduction have relatively large wavefunction coefficients; once identified, they persist across iterations, reducing the rate of new carryover additions. This observation confirms that the carryover mechanism operates as intended for this problem size.

Figure~\ref{fig:dashboard-sqd}(c) shows that the parameter convergence metric diverges in both executions, indicating that trial populations remain widely spread. While we do not attempt to explain this behavior scientifically in this work, such observability provides actionable insights: for example, divergence may suggest that hyperparameters like $F = 0.6$ in Eq.~\eqref{eq:mutant} may require adjustment. Although tuning still requires additional experiments, telemetry helps identify potential causality between parameter settings and workflow behavior, reducing reliance on blind or ad hoc parameter searches.

Figure~\ref{fig:dashboard-sqd}(d) reports the Hamming distance, which converges to approximately 4.2. This indicates that the final wavefunction differs from the RHF reference by an average of 4.2 orbital occupations. While assessing the physical validity of this value is beyond the scope of this work, future studies could compare this metric across different molecular systems using the same observability framework.

Figure~\ref{fig:dashboard-sqd}(e) shows that the sample preservation ratio remains around 0.01\%. This result suggests that noise at the given circuit depth is substantial, even with general-purpose error mitigation, and that configuration recovery significantly alters the sampled distribution. Such behavior may contribute to non-reproducibility, as noise-induced variations could influence the discovery of favorable determinants.

Finally, Figure~\ref{fig:dashboard-sqd}(f) displays orbital occupancy trends across iterations. For the [4Fe-4S] cluster with $N_e = 27$, orbital index 26 corresponds to the Fermi level in the RHF state. We focus on the lowest unoccupied molecular orbital (LUMO), where occupancy increases as optimization progresses, indicating contributions from higher-level orbitals to the ground-state energy. Although only the first execution is shown, the second exhibited a similar trend.

\subsection{Performance analysis}

System performance metrics are critical not only for administrators but also for algorithm developers, as poorly optimized workflows can rapidly consume expensive resources. Figures~\ref{fig:dashboard-performance}(c) and (f) show that the wall-clock time for a single QPU job is approximately 5 minutes, while an HPC job takes around 10 minutes. Executing such workflows in a tightly coupled system or under resource reservations would result in significant idle time because QPU and HPC tasks are inherently dependent and cannot run concurrently. Previous work~\cite{shirakawa2025closedloopcalculationselectronicstructure} mitigated this by running two optimizations in parallel, but this approach requires complex workflow control and careful hyperparameter tuning to maintain efficiency.

Our loosely coupled design avoids idle time but introduces variability in queuing delays, as shown in Figures~\ref{fig:dashboard-performance}(b) and (e). While these delays are often shorter than job execution time, they can occasionally dominate workflow turnaround due to scheduler policies beyond user control. For time-critical applications, tightly coupled systems may still be preferable.

Cost management is another important consideration. Figures~\ref{fig:dashboard-performance}(a) and (d) indicate that a single execution consumes 19.6\% of the QPU allocation and 7.7\% of the HPC allocation, limiting the number of feasible runs to approximately five under current quotas. This constraint underscores the need for telemetry-driven optimization to minimize wasted resources during hyperparameter tuning.

Figures~\ref{fig:dashboard-performance}(g) and (h) show resource utilization on Miyabi-C compute nodes. Despite access to roughly 7.92 TB of memory and 7168 cores, only 28–35\% of cores were active during execution. This suggests potential improvements in job configuration, such as adjusting MPI process counts, OpenMP threading, or solver parameters. Memory usage also indicates that the subspace dimension $d$ could be increased to include more determinants for improved energy estimates without exceeding system limits.

Figure~\ref{fig:dashboard-performance}(i) highlights task-level execution times. As expected, \texttt{run\_primitive} and \texttt{solve\_eigenstate} dominate, consuming 48.4k and 97k seconds respectively. These tasks are primarily I/O-bound because they involve submitting jobs to QPU and HPC systems, making queuing delays a major contributor to overall workflow time. This observation reinforces the importance of advanced scheduling strategies to improve throughput. Telemetry also reveals smaller tasks like \texttt{recover\_configurations} (4070 seconds) that could be optimized. Since the task executes a Python function~\cite{qiskit-addon-sqd} directly on the \texttt{cli} machine, one could offload to Miyabi-C or rewrite in a compiled language. While such changes would have a minor impact compared to queuing delays, they illustrate how observability can uncover incremental improvements.

Overall, these observations demonstrate how telemetry enables systematic identification of performance bottlenecks and resource inefficiencies, guiding both workflow design and system-level optimization.

\section{Conclusion\label{sec:conclusion}}

In this work, we introduced the concept of application-level telemetry for quantum-centric supercomputing (QCSC) workflows and formalized it through the workflow metrics pyramid. This conceptual framework highlights the need for observability beyond system-centric metrics, extending to domain-level insights that inform algorithm design and scientific interpretation.

Building on this concept, we proposed an observability architecture that decouples the computing and observability platforms, enabling persistent telemetry collection, structured storage, and post-hoc analysis without interfering with primary workload execution. We demonstrated a reference implementation on the Miyabi supercomputer and IBM Quantum systems, showcasing its utility in analyzing solver behavior and performance characteristics in a closed-loop SQD workflow.

A key feature of our design is that telemetry data is persistently stored in the observability server. This allows new metrics to be introduced without re-executing workflows, requiring only modifications to the ETL pipeline and dashboard. Such flexibility is critical for iterative research, where new hypotheses often emerge after initial experiments.

While our implementation primarily focused on L2–L4 telemetry, L0 and L1 metrics remain unexplored due to limited access to administrator-level APIs. Incorporating these metrics could provide valuable correlations—for example, between system resource contention and observed variability in job wall-clock times, or between hardware drift (e.g., cryostat base temperature) and algorithmic anomalies. Such integration would strengthen the observability framework by linking infrastructure-level dynamics with application-level outcomes.

Another limitation is the queuing delay observed in loosely coupled execution. Although this design avoids idle time across QPU and HPC resources, it introduces variability in workflow turnaround time. Future work could explore co-scheduling strategies, leveraging emerging features in job schedulers such as SLURM’s quantum resource reservation~\cite{bacher2025quantumresourcesresourcemanagement, esposito2025slurmheterogeneousjobshybrid}, to optimize joint allocation of classical and quantum resources.

Finally, while we demonstrated the architecture using SQD with differential evolution, the proposed framework is broadly applicable to other hybrid algorithms, including variational quantum eigensolvers, quantum machine learning workflows, and sample-based Krylov methods to name a few~\cite{yu2025quantumcentricalgorithmsamplebasedkrylov, yoshioka2024diagonalization, nakagawa2024adapt, piccinelli2025quantumchemistryprovableconvergence, zhang2025random, sakka2025automating}. By extending telemetry coverage and integrating advanced scheduling mechanisms, this approach can evolve into a comprehensive tool for infrastructure-aware algorithm development and performance optimization.

\section*{Author Contributions}
N.K conceived the project, designed the architecture, implemented and executed the closed-loop SQD workflow, and wrote the manuscript.
Y.M implemented the data pipeline, designed the observability dashboard, conducted analysis, and contributed to manuscript preparation.
H.T setup the experiment environment on the \texttt{serv} machine.
Y.K supervised design of L4 metrics.
H.H provided strategic guidance and revised the manuscript.
All authors discussed the results and approved the final version.

\section*{Acknowledgement}
We acknowledge fruitful discussions with Toshinari Itoko.
We acknowledge people creating and supporting the \texttt{ibm\_kingston} system on which all data presented here was taken.
This research was conducted using the Myabi System at the Joint Center for Advanced High Performance Computing (JCAHPC).

\bibliographystyle{myapsrev4-1}
\bibliography{references}

\end{document}